\documentclass[runningheads,envcountsame]{llncs}
\usepackage{microtype}
\usepackage{amsmath,amssymb}
\usepackage{proof}
\usepackage{stmaryrd}
\usepackage{latexsym}
\usepackage{url}
\usepackage{mycommands}
\usepackage{color}
\usepackage{xspace}
\usepackage{relsize}
\usepackage{times}
\usepackage{tikz}
\usepackage[all]{xy}
\RequirePackage{txfonts}
\usepackage[show]{ed}
\usepackage{hyperref}
\usepackage{graphicx}
\usepackage{wrapfig}
\usepackage{wasysym}
\usepackage{mathtools}
\usepackage{cleveref}
\usepackage{float}
\usepackage{caption}
\usepackage{subcaption}
\usepackage{mathdots}
\usepackage{comment}

\title{Multi-modalities and non-commutativity/associativity in functorial linear logic: a case study}
\titlerunning{Non-commutative/associative multi-modal functorial linear logic}

\author{
Carlos Olarte\inst{1} \and
Elaine Pimentel\inst{2}}
\institute{LIPN, Universit\'{e} Sorbonne Paris Nord, France \and
Dept. of Computer Science, University College London, UK }

\begin{document}
\maketitle



\section{Introduction}

Similar to modal connectives, the exponential $\bang$ in intuitionistic linear logic ($\ILL$) is not {\em canonical}~\cite{danos93kgc}, in the sense that if  $i\not= j$ then
$\nbang{i}F\not\equiv\nbang{j}F$.
Intuitively, this means that we can mark the exponential with {\em labels} taken from a set $I$ organized in a pre-order $\preceq$ (\ie, reflexive and transitive), obtaining (possibly infinitely-many) exponentials ($\nbang{i}$
for $i\in I$).
Also as in multi-modal systems, the pre-order  determines the provability relation: 
for a general formula $F$, $\nbang{b}F$ {\em implies} $\nbang{a}F$ iff $a \preceq b$.

The algebraic structure of subexponentials, combined with their intrinsic structural property  allow for the proposal of rich linear logic based frameworks. This opened a venue for proposing different multi-modal substructural logical systems, that encountered  a number of different applications.

There are, however, two main differences between multi-modalities in normal modal logics and  subexponentials in linear logic.

\noindent\textbf{i. substructural behaviour.} Subexponentials carry the possibility of 
having different structural behaviors;

\noindent\textbf{ii. nature of modalities.}
Normal modal logics start from the weakest version, assuming only axiom $\K$,
then extensions are considered, by adding other axioms. Exponentials in linear
logic
``take for granted'' the behaviors expressed by axioms
$\T$ and $\4$. 

Regarding (i), originally~\cite{nigam10jar}, subexponentials could assume only weakening and contraction axioms, while in~\cite{DBLP:conf/cade/KanovichKNS18,DBLP:journals/mscs/KanovichKNS19}, non-commutative systems allowing commutative subexponentials were presented
\[ \C:\;\; \nbang{i} F \limp \nbang{i} F
     \otimes \nbang{i} F \qquad \W:\;\; \nbang{i} F \limp \one  
\qquad \mathsf{E}:\;\; (\nbang{i} F)\otimes G \equiv G\otimes(\nbang{i} F) \]
Finally, in~\cite{DBLP:conf/cade/BlaisdellKKPS22}  associativity was added to the scene, where exponentials could 
assume or not the axioms   
\[  \mathsf{A1}:\;\; \nbang{i} F\otimes(G\otimes H) \limp (\nbang{i} F\otimes G)\otimes H
     \qquad \mathsf{A2}:\;\; (G\otimes H)\otimes \nbang{i} F \limp G\otimes (H\otimes\nbang{i} F) \]

Concerning (ii),  Guerrini et al.~\cite{DBLP:journals/igpl/GuerriniMM98} unified the modal and
$\LL$ approaches, with the exponentials assuming only the linear
version of $\K$, with the possibility of adding modal  extensions to it. This discussion was brought to multi-modal case
in~\cite{DBLP:conf/lpar/LellmannOP17}, where
subexponentials consider not only the structural axioms 
 for contraction and weakening, but also the subexponential version of axioms
     $\{\K,\4,\D,\T\}$: \[ \K:\;\; \nbang{i} (F \limp G) \limp \nbang{i} F \limp
         \nbang{i} G \qquad \D:\;\; \nbang{i} F \limp \nquest{i} F \qquad \T:\;\;
     \nbang{i} F \limp F \qquad \4\:;\; \nbang{i} F \limp \nbang{i}\nbang{i} F
 \] 
 
 In this work, we intend to join these two studies. This means that $\nquest{i}$ can behave classically or not, model associative and commutative systems or not,
 but also with exponential behaviors different from those in $\LL$.  Hence, by
 assigning different  modal axioms  one obtains, in a modular way, a class of
 different substructural modal logics.  
 
\section{Non-commutative, non-associative linear logic ($\acLL$)}
In the following, we will briefly describe the system $\acLL$. For more details, please refer to~\cite{DBLP:conf/cade/BlaisdellKKPS22}.

\noindent
{\em Connectives.}
First of all, non-commutativity implies that the left residual $\bla$ should be added to the language of (propositional intuitionistic) linear logic with subexponentials ($\SELL$), which contains the connectives $\tensor,\with,\oplus,\one,\zero,\limp,\nbang{l},\nquest{l}$. The subexponentials are organized as follows. 
\begin{definition}[Signature]\label{def:sdmls}
Let $\mathcal{A}=\{\C,\W,\mathsf{A1},\mathsf{A2},\mathsf{E}\}$. A 
{\em subexponential signature}  is given
 by a triple $\Sigma=(I,\cless,f)$, where $I$ is a set of indices, $(I,\cless)$ is a pre-order, and $f$ is a mapping from $I$ to 
  $2^{\mathcal{A}}$. Finally, every signature $\Sigma$ is assumed to be  upwardly closed w.r.t. $\preceq$, that is, if $i\preceq j$ then $f(i)\subseteq f(j)$ for all $i,j\in I$.
\end{definition}

\noindent
{\em Contexts and rules.}
Losing commutativity implies that contexts should be handled by lists instead of multisets, and the lack of associativity forces tree-shaped contexts. 
Rules then act deeply in these structures. For example, the promotion and exchange rules are 
\[\infer[\nbang{i} R]{\Gamma\seq \nbang{i}F}{\Gamma^{\upset{i}}\seq F}\qquad \infer[\mathsf{E}1]{\Rx{(\nbang{e}\Delta_1,\Delta_2)}\seq G}{\Rx{(\Delta_2,\nbang{e}\Delta_1)}\seq G}
\qquad
\infer[\mathsf{E}2]{\Rx{(\Delta_1,\nbang{e}\Delta_2)}\seq G}{\Rx{(\nbang{e}\Delta_2,\Delta_1)}\seq G}
\]
where $\upset{i}$ denotes the \emph{upset} of the index $i$, \ie, the set $\{j \in I : i \cless j\}$.   
\begin{example}\label{ex:preceq}
Let $\Gamma=(\nbang{i}A,(\nbang{j}B,\nbang{k}C))$ be represented below left, $i\preceq j$ but $i\not\preceq k$, and $\W\in f(k)$. Then $\Gamma^{\upset{i}}=(\nbang{i}A,\nbang{j}B)$ is depicted below right
{\small
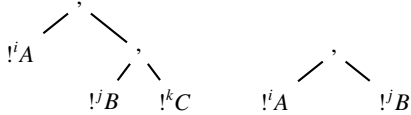
\begin{center}
\begin{tikzpicture}[level distance=2em, 
			level 1/.style={sibling distance=5em},
			level 2/.style={sibling distance=3em},
			every node/.style = {align=center}]]
			\node {,}
			child[thick] { node {$\nbang{i}A$}}
			child[thick] { node {,}
				child[thick] { node {$\nbang{j}B$}}
				child[thick] {node {$\nbang{k}C$}}};
		\end{tikzpicture}
 \qquad
\begin{tikzpicture}[level distance=2em, 
			level 1/.style={sibling distance=5em},
			level 2/.style={sibling distance=3em},
			every node/.style = {align=center}]]
			\node {,}
			child[thick]{node {$\nbang{i}A$}}
			child[thick] { node {$\nbang{j}B$}};
		\end{tikzpicture}
\end{center}
}
Observe that, if $\W\notin f(k)$, then $\Gamma^{\upset{i}}$ cannot be built. In this case, any derivation of  $\Gamma\seq\nbang{i}(A\tensor B)$ cannot start with an application of the promotion rule $\nbang{i} R$ (similarly to how promotion in $\ILL$ cannot be applied in the presence of non-classical contexts). In this case, if $A,B$ are atomic, this  sequent would not be provable.
\end{example}

\section{Linear nested systems}
In this section we  present a brief introduction to 
linear nested systems. For further details, refer to  (\cite{DBLP:conf/tableaux/Lellmann15,lellmann19tocl}).

One of the main problems of using sequent systems as a framework is that sequents are often not adequate for expressing modal behaviors. 
In order to propose a better formulation, 
we need a tighter control of formulas in the context, something that sequents do not provide. Hence the need for extending the notion of sequent systems.
\begin{definition}[LNS]
The set $\LNS$ of \emph{linear nested sequents} is given recursively by:

\noindent (i) if  $\Gamma\seq \Delta$ is a sequent then $\Gamma\seq \Delta\in\LNS$

\noindent (ii)  if $\Gamma\seq \Delta$ is a sequent and $\mathcal{G}\in\LNS$
    then $\Gamma\seq\Delta\lns\mathcal{G}\in\LNS$.

We call each sequent
in a linear nested sequent a {\em component} and slightly abuse
notation, abbreviating ``linear nested sequent'' to $\LNS$.  
    We shall denote by $\LNS_{\mathcal{L}}$ a linear nested sequent system
  for a logic $\mathcal{L}$.
\end{definition}
In words,  a linear nested sequent is simply a finite list of sequents that matches exactly the \emph{history} of a backward proof search in an ordinary sequent
calculus~(\cite{DBLP:conf/tableaux/Lellmann15}). 

A further advantage of this framework is that it is often possible to restrict the list of sequents in a $\LNS$ to the last 2 components, that we call {\em active}.

\begin{definition}[End-active]\label{end-active}\label{def:end-active}
  An application of a linear nested sequent rule is \emph{end-active}
  if the rightmost components of the premises are active and the only
  active components (in premise and conclusion) are the two rightmost
  ones. The \emph{end-active variant} of a $\LNS$ calculus is the
  calculus with the rules restricted to end-active applications.
\end{definition}

\subsection{$\LNS$ for multi-modal $\LL$}
In the quest for locality,  \cite{DBLP:journals/igpl/GuerriniMM98}  
proposed 2-sequents systems  for $\LL$ variants, with separate rules for the exponentials. In~\cite{DBLP:conf/lpar/LellmannOP17} this work was revisited, establishing a lighter notation and extending the discussion to multi-modalities. 

$\LNS_\LL$ (\cite{DBLP:conf/lpar/LellmannOP17}) is  an end-active, linear nested system for linear logic.
In this system, the promotion rule is split into the following local rules:
%
\[
\infer[\bang R]{\Escr \lns  \Gamma\seq !F}{
\vdash \Gamma \seq \lns \seq F
}
\qquad
\infer[\bang L]{ \Gamma, \bang F\seq H \lns \Delta\seq G}{\Gamma\seq H \lns \Delta, \bang F\seq G}
\]
Observe that no checking must be done in the context in order to apply the $\bang L$ rule: The only checking is in the $\bang R$ rule, where $\Escr$ should be the empty sequent or an empty list of components. 

 More precisely, applying the $\bang$ rule enables the creation of the {\em future history}, in which the banged formula should be proved. 
The
intended interpretation of a $\LNS$ in $\LL$ is
$$
\begin{array}{lcl}
\iota( \Gamma\seq F) & \defs& \bigotimes \Gamma\limp F\\
\iota(\Gamma \seq F  \lns \mathcal{H}) & \defs& \bigotimes  (\Gamma\limp F)  \lpar \; \bang \iota(\mathcal{H})
\end{array}
$$

The notion of signatures is then enhanced.  We say that $\Sigma=(I,\cless,f)$ is a {\em functorial signature}
if it is defined over $\mathcal{A}=\{\C,\W,\mathsf{A1},\mathsf{A2},\mathsf{E},\K,\4,\T,\D\}$ and $\K\in f(i),\;\forall i\in I$. If, moreover, $\mathsf{A1},\mathsf{A2}\in f(i),\;\forall i\in I$, then the functorial signature is called {\em associative}.
 
\subsubsection{The associative case.} In the presence of associativity, contexts are lists of formulas, and the tree structure is not needed. 
Given an associative signature $\Sigma$,
 we introduce nesting operators $\lns[i]$ and their
unfinished versions $\lnse[i]$ for every $i
\in I$, and change the interpretation so that they are interpreted by
the corresponding modality:
\begin{align*}
  \iota(\Gamma \seq F) & \defs  \bigotimes \Gamma\limp F \\
  \iota(\Gamma \seq F  \lns[i] \mathcal{H}) & \defs \iota(\Gamma\seq F \lnse[i]
                                           \mathcal{H}) \defs (\bigotimes \Gamma\limp F)\;
                                           \lpar \nbang{i} \,\iota(\mathcal{H})\end{align*}
The operators $\lnse[i]$  
indicate that the standard sequent rule for the modality indexed by $i$
has been partially processed as shown below. 

The exponential rules for the 
end-active linear nested system $\LNS_{(I,\cless,F)}$ is given by   the following rules
\[
  \infer[\nbang{i}_{\mathsf{k}}\;(\text{for }j\in\upset{i})]{\Gamma,\nbang{j} F\seq G\lnse[i]\Delta\seq H}{\Gamma\seq G\lnse[i] F,\Delta\seq H}\qquad 
   \infer[\nbang{i}]{\Gamma\seq \nbang{i} F}{\Gamma\seq \lnse[i] \seq F}\qquad
   \infer[\mathsf{r}]{\mathcal{E}\lnse[i]\Gamma\seq F}{\Gamma\seq F}
\]
\[
       \infer[\nbang{i}_{\4}\;(\text{for }j \in {\upset[\4]{i}})]{\Gamma,\nbang{j} F\seq G\lnse[i]\Delta\seq H}{\Gamma\seq G\lnse[i] \nbang{j} F,\Delta\seq H}
        \quad 
    \infer[\nbang{i}_{\mathsf{d}}\;(\text{for }\D \in F(i))]{\Gamma,\nbang{i} F\seq G}{\Gamma\seq G\lnse[i] F\seq}
\]
\[
\infer[\nbang{i}_{\mathsf{t}}\;(\text{for }\T \in F(i))]{\Rx{\nbang{i}F}\seq G }{\Rx{F}\seq G}
\]
Note that the order of formulas in the context is respected in the application of the rules.

\subsection{The non-associative case}
Since $\LNS$ do not preserve tree structures, the system  just proposed is not adequate for the non-associative setting. In this case, we propose a skeleton that  keeps track of the shape of the structure as follows. 

Given a structure $\Gamma$, we represent by $\Gamma^{\circ}$ its underlying tree structure, where the leaves of $\Gamma$ are substituted by empty contexts $\{\;\}$ (holes). $\Gamma^{\circ}\{\}$ will then represent $\Gamma^{\circ}$ with a specific position highlighted, and the usual context substitution can be applied.  That is, given an underlying structure $\Rxc$, we write $\Rxc{F}$  for the context where some specific hole $\Rxc{}$ has been replaced by $F$. If $\Rxc=\EMP$ the  hole is removed.

The $\LNS$ rules are then adapted from the rules of the associative case, for example
\[
  \infer[\nbang{i}_{\mathsf{k}}\;(\text{for }j\in\upset{i})]{\Rx{\nbang{j} F}\seq G\lnse[i]\Rxc{}\seq H}{\Rx{}\seq G\lnse[i] \Rxc{F}\seq H}\qquad 
   \infer[\nbang{i}]{\Gamma\seq \nbang{i} F}{\Gamma\seq \lnse[i] \;\Gamma^{\circ}\seq F}
   \qquad
   \infer[\mathsf{r}]{\mathcal{E}\lnse[i]\Gamma^{\circ}\seq F}{\Gamma\seq F}
   \]
   \[
    \infer[\nbang{i}_{\mathsf{d}}\;(\text{for }\D \in F(i))]{\Rx{\nbang{i} F}\seq G}{\Gamma\seq G\lnse[i] \Rxc{F}\seq}
\]

\begin{example}
Let $\Gamma$ be as in Example~\ref{ex:preceq}, and suppose that $j \in\upset[\4]{i}, \4\notin f(i)$ and $i\not\preceq k$. Below are possible configurations of $\Gamma ^{\preceq i}$ after applying a $\K\4_i$ rule 
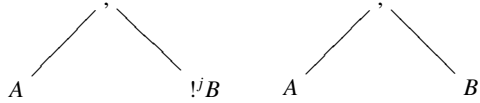
\small{\[
\xymatrix{
				& , & \\
				A \ar@{-}[ur] & &  \nbang{j}B \ar@{-}[ul]  
			} \qquad
\xymatrix{
				& , & \\
				A \ar@{-}[ur] & &  B \ar@{-}[ul]  
			}
\]}
\end{example}

\section{Conclusion}
In this ongoing work, we propose a modular system for dealing with multi-modal $\LL$ systems, where substructural and modal axioms are taken into account.

The motivation for that is twofold: (1) to expand the role of $\LL$ as a framework for reasoning about systems, as in~\cite{DBLP:journals/tcs/MillerP13,nigam14jlc,DBLP:journals/mscs/XavierOP22} and, mainly, (2) extending the computational interpretation of subexponentials~\cite{DBLP:journals/tcs/NigamOP17,DBLP:conf/fscd/PimentelON21} also to the non-associative/commutative case. We intend to pursue these directions in the near future.

%
%


\end{document}